\documentclass[
reprint,
amsmath,
amssymb,
aip,
pop
]{revtex4-2}

\usepackage[USenglish]{babel}
\usepackage{graphicx}
\usepackage{booktabs}
\usepackage{url}
\usepackage{tikz,circuitikz}
\usetikzlibrary{decorations.pathreplacing,arrows.meta,calc,patterns,positioning}
\usepackage[deletedmarkup=xout]{changes}
\definechangesauthor[color=orange]{VM}

\definechangesauthor[color=magenta]{TMu}

\begin{document}
	
	\title{A Nonlinear Two-Sheath Circuit Model for Low-Pressure Symmetric and Asymmetric Capacitively Coupled Radio-Frequency Plasmas}
	\author{Katharina Noesges}
	\affiliation{Chair of Applied Electrodynamics and Plasma Technology, Faculty of Electrical Engineering and Information Technology, Ruhr University Bochum, 44780 Bochum, Germany}
	\author{Tim Bolles}
	\affiliation{Chair of Applied Electrodynamics and Plasma Technology, Faculty of Electrical Engineering and Information Technology, Ruhr University Bochum, 44780 Bochum, Germany}
	\author{Máté Vass}
	\affiliation{Chair of Applied Electrodynamics and Plasma Technology, Faculty of Electrical Engineering and Information Technology, Ruhr University Bochum, 44780 Bochum, Germany}	\author{Ihor Korolov}
	\affiliation{Chair of Applied Electrodynamics and Plasma Technology, Faculty of Electrical Engineering and Information Technology, Ruhr University Bochum, 44780 Bochum, Germany}
	\author{Thomas Mussenbrock}
	\email{Thomas.Mussenbrock@rub.de}
	\affiliation{Chair of Applied Electrodynamics and Plasma Technology, Faculty of Electrical Engineering and Information Technology, Ruhr University Bochum, 44780 Bochum, Germany}
%	\date{\today}
\date{October 1, 2026}
	
	\begin{abstract}
		We develop a nonlinear self-consistent two-sheath circuit model for low-pressure capacitively coupled plasmas that applies to geometrically symmetric as well as asymmetric discharges. The quasineutral plasma bulk is represented by an inductive--resistive element and coupled to stationary particle and electron-energy balances. Both boundary sheaths are treated dynamically using a lncosh sheath charge--voltage model with bounded differential elastance, based on a Riccati closure for the differential sheath width. The model recovers the quadratic depletion-sheath relation in the small-charge limit, while the characteristic sheath scales are determined from the RF-averaged sheath voltages using a collisionless Child--Langmuir/Bohm closure. The resulting four-variable RF subsystem contains the two sheath charges, the blocking-capacitor voltage, and the discharge current. In the symmetric monofrequent limit, the two sheath nonlinearities compensate strongly and the dc self-bias vanishes, whereas geometrical or electrical asymmetry breaks this compensation and enhances harmonic generation and plasma-series-resonance oscillations.
	\end{abstract}
	
	\maketitle
	
	\section{Introduction}
	
	Capacitively coupled radio-frequency (CCRF) discharges are intrinsically nonlinear electrical systems. A substantial fraction of the applied RF voltage is supported by the plasma boundary sheaths, whose nonlinear charge--voltage response has been treated in analytical sheath models and reduced reactor descriptions for several decades \cite{Metze1986,Lieberman1988,GodyakSternberg1990,Klick1996,LiebermanLichtenberg2005}. The quasineutral bulk, in turn, exhibits an inertial and dissipative electron response, so that the sheath--bulk system forms a driven nonlinear oscillator. The combination of sheath capacitance and bulk inertia gives rise to plasma--sheath series resonances, identified experimentally and interpreted with equivalent circuits in early work by Godyak and Popov, Annaratone, Allen, Ku, and co-workers, and later in simulations of resonant bounded plasmas \cite{GodyakPopov1979,AnnaratoneAllenKu1995,KuAnnaratoneAllen1998a,KuAnnaratoneAllen1998b,CooperbergBirdsall1998,QiuBowersBirdsall2003}. In a nonlinear discharge, harmonics generated by the sheaths can couple to this resonant response \cite{KlickRehakKammeyer1997,MussenbrockBrinkmann2006APL}. Geometrical asymmetry breaks the equivalence of the two boundary sheaths and thereby prevents cancellation of their nonlinear voltage contributions, providing a direct route to plasma series resonance (PSR) excitation \cite{CzarnetzkiEtAl2006POP,Lieberman2008POP}. Experiments and kinetic simulations have, however, shown that these oscillations are not restricted to strongly geometrically asymmetric discharges \cite{SchulzeEtAl2007JPCS,DonkoEtAl2009APL,SchuengelEtAl2015POP,WilczekEtAl2016POP}.
	
	Reduced lumped-element and global models are useful because they retain the sheath--bulk interaction with only a few dynamical variables. Besides early nonlinear circuit and global descriptions \cite{Metze1986,MussenbrockZieglerBrinkmann2006POP}, more recent global models have treated the nonlinear current and dc self-bias of single- and multi-frequency CCPs with explicit sheath dynamics \cite{SaikiaEtAl2018AIPAdv,SaikiaEtAl2018JAP}. A recent self-consistent circuit model combined one dominant dynamic sheath, a blocking capacitor, and an inertial--resistive plasma bulk with stationary particle and electron-energy balances \cite{Mussenbrock2026Minimal}. That construction is particularly suited to strongly asymmetric CCPs, where one boundary sheath carries most of the RF voltage modulation. Extending this approach to weakly asymmetric and symmetric discharges therefore requires both boundary sheaths to be treated dynamically.
	
	A direct extension of a one-sheath quadratic model to two identical quadratic sheaths has a known structural limitation. Equivalent-circuit analyses of symmetric and weakly asymmetric CCPs have shown that the quadratic sheath nonlinearities can compensate, while cubic sheath terms and a time-dependent bulk electron plasma frequency can restore the nonlinearity required for self-excited PSR oscillations \cite{SchuengelEtAl2015POP}.
	
	Neglecting the comparatively small net conduction charge transferred over an RF period, the two sheath charges may be written as
	$Q_{\mathrm E}=\overline Q+q$
	and
	$Q_{\mathrm G}=\overline Q-q$,
	where $\overline Q = (Q_{\rm E}+Q_{\rm G})/2$ denotes the mean sheath charge and $q$ the RF charge excursion. For identical sheaths with charge--voltage characteristic $V_{\mathrm S}(Q)$, a quadratic law $V_{\mathrm S}(Q)=aQ^2$ gives
	\begin{gather}
		V_{\mathrm S}(\overline Q+q)-V_{\mathrm S}(\overline Q-q)
		=4a\overline Q q,
	\end{gather}
	which is exactly linear in the RF charge excursion. The same sheath-symmetry considerations form the basis of the electrical asymmetry effect (EAE): an even harmonic in the applied waveform breaks the electrical symmetry of the two sheaths even for a geometrically symmetric reactor \cite{HeilEtAl2008JPD,DonkoEtAl2009JPD,CzarnetzkiEtAl2009JPCS,SchulzeEtAl2009JPD}. Accordingly, a two-sheath circuit model applicable across symmetric and asymmetric configurations must retain higher-order contributions to the sheath charge--voltage characteristic.
	
	To address this limitation, the present model builds on the self-consistent circuit and global framework of Ref.~\cite{Mussenbrock2026Minimal}, retaining its stationary argon closure, inertial--resistive plasma bulk, blocking capacitor, power balance, and numerical coupling. The principal extension is the dynamic treatment of both boundary sheaths using a lncosh sheath charge--voltage model with bounded differential elastance (inverse differential capacitance). A function of this form was already used by Ziegler et al. as an asymptotically linear representation of an experimentally reconstructed sheath characteristic \cite{ZieglerMussenbrockBrinkmann2008PSST}. Here, its Riccati structure is connected explicitly to the differential sheath width. The two characteristic sheath scales are determined independently from their RF-averaged voltage drops. A general multi-harmonic source then allows geometrical and electrical symmetry breaking to be studied within the same four-variable circuit model.
	
	\section{Nonlinear two-sheath circuit model}
	
	\subsection{Scale separation and equivalent geometry}
	
	We consider a low-pressure CCRF discharge for which
	\begin{gather}
		\lambda_{\mathrm D}\ll l_{\mathrm s}<d_{\mathrm{gap}}\ll\lambda_{\mathrm{EM}},
	\end{gather}
	and
	\begin{gather}
		\omega_{\mathrm{pi}}\ll\omega_{\mathrm{RF}}\ll\omega_{\mathrm{pe}}.
	\end{gather}
	Here, \(\lambda_{\mathrm D}\) is the Debye length, \(l_{\mathrm s}\) a characteristic sheath thickness, \(d_{\mathrm{gap}}\) the electrode separation, and \(\lambda_{\mathrm{EM}}\) the electromagnetic wavelength associated with the RF excitation. The quantities \(\omega_{\mathrm{pi}}\), \(\omega_{\mathrm{RF}}\), and \(\omega_{\mathrm{pe}}\) denote the ion plasma, driving, and electron plasma frequencies, respectively.
	The non-neutral sheaths are therefore separated from the quasineutral bulk, ion motion is slow on the RF timescale, while electrons can respond collectively on this timescale. Electron inertia is retained in the bulk momentum balance to capture its inductive RF response \cite{LiebermanLichtenberg2005}. The model is restricted to the electroquasistatic regime; large-area standing-wave and skin-effect phenomena are therefore not considered \cite{LiebermanBoothChabert2002,Chabert2007,LeeGravesLieberman2008}. The macroscopic plasma state, represented by the electron density $n_{\mathrm e}$ and temperature $T_{\mathrm e}$, is assumed constant during one RF period and is determined from stationary particle and energy balances.
	
	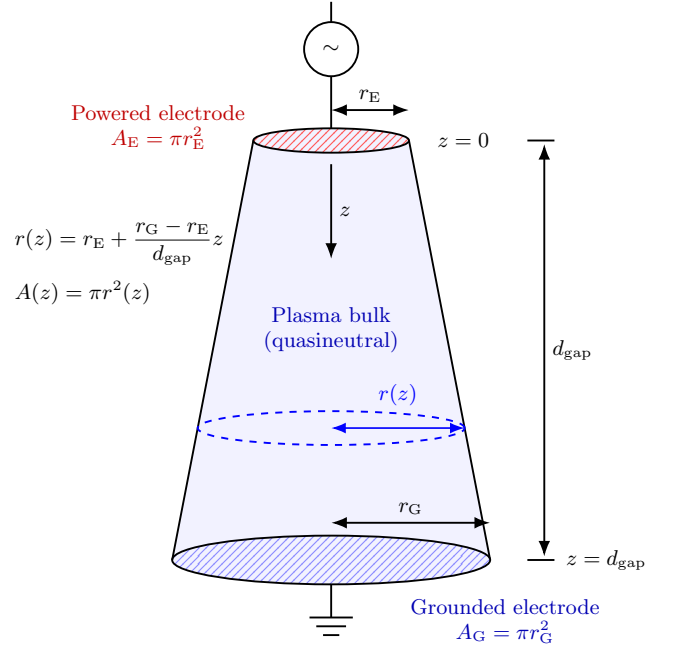
\begin{figure}[t]
		\centering
		\resizebox{\columnwidth}{!}{%
			
			\begin{tikzpicture}[
				>=Latex,
				line width=0.8pt,
				font=\small
				]
				
				% ------------------------------------------------------------
				% Geometry
				% ------------------------------------------------------------
				
				\def\rE{1.15}
				\def\rG{2.35}
				\def\h{6.2}
				
				% Electrode vertical positions
				\coordinate (ET) at (0,0);
				\coordinate (EG) at (0,-\h);
				
				% ------------------------------------------------------------
				% Plasma bulk: conical frustum
				% ------------------------------------------------------------
				
				\fill[blue!5]
				(-\rE,0)
				--
				(-\rG,-\h)
				arc[start angle=180,end angle=360,x radius=\rG,y radius=0.36]
				--
				(\rE,0)
				arc[start angle=0,end angle=180,x radius=\rE,y radius=0.18]
				-- cycle;
				
				% Side walls
				\draw
				(-\rE,0)
				--
				(-\rG,-\h);
				
				\draw
				(\rE,0)
				--
				(\rG,-\h);
				
				% ------------------------------------------------------------
				% Powered electrode
				% ------------------------------------------------------------
				
				\fill[
				red!25,
				pattern=north east lines,
				pattern color=red!65
				]
				(0,0)
				ellipse[x radius=\rE,y radius=0.18];
				
				\draw
				(0,0)
				ellipse[x radius=\rE,y radius=0.18];
				
				% ------------------------------------------------------------
				% Grounded electrode
				% ------------------------------------------------------------
				
				\fill[
				blue!18,
				pattern=north east lines,
				pattern color=blue!45
				]
				(0,-\h)
				ellipse[x radius=\rG,y radius=0.36];
				
				\draw
				(0,-\h)
				ellipse[x radius=\rG,y radius=0.36];
				
				% ------------------------------------------------------------
				% RF source
				% ------------------------------------------------------------
				
				\draw (0,0.18) -- (0,0.95);
				
				\draw (0,1.35) circle[radius=0.40];
				
				\node at (0,1.35) {$\sim$};
				
				\draw (0,1.75) -- (0,2.05);
				
				% ------------------------------------------------------------
				% Ground
				% ------------------------------------------------------------
				
				\draw (0,-\h-0.36) -- (0,-\h-0.85);
				
				\draw (-0.32,-\h-0.85) -- (0.32,-\h-0.85);
				\draw (-0.22,-\h-0.98) -- (0.22,-\h-0.98);
				\draw (-0.12,-\h-1.11) -- (0.12,-\h-1.11);
				
				% ------------------------------------------------------------
				% Axial coordinate
				% ------------------------------------------------------------
				
				\draw[->]
				(0,-0.35)
				--
				(0,-1.75);
				
				\node[right] at (0,-1.05) {$z$};
				
				% ------------------------------------------------------------
				% Radius r_E
				% ------------------------------------------------------------
				
				\draw[<->]
				(0,0.45)
				--
				(\rE,0.45);
				
				\node[above] at ({0.5*\rE},0.45) {$r_{\mathrm E}$};
				
				% ------------------------------------------------------------
				% Radius r_G
				% ------------------------------------------------------------
				
				\draw[<->]
				(0,-\h+0.55)
				--
				(\rG,-\h+0.55);
				
				\node[above] at ({0.5*\rG},-\h+0.55) {$r_{\mathrm G}$};
				
				% ------------------------------------------------------------
				% Local cross section
				% ------------------------------------------------------------
				
				\def\zloc{-4.25}
				\pgfmathsetmacro{\rloc}{
					\rE + (\rG-\rE)*abs(\zloc)/\h
				}
				
				\draw[
				blue,
				dashed
				]
				(0,\zloc)
				ellipse[
				x radius=\rloc,
				y radius=0.25
				];
				
				\draw[
				blue,
				<->,
				line width=0.7pt
				]
				(0,\zloc)
				--
				(\rloc,\zloc);
				
				\node[
				blue,
				above
				] at ({0.5*\rloc},0.95*\zloc)
				{$r(z)$};
				
				% ------------------------------------------------------------
				% Gap distance
				% ------------------------------------------------------------
				
				\draw
				(\rG+0.55,0)
				--
				(\rG+0.95,0);
				
				\draw
				(\rG+0.55,-\h)
				--
				(\rG+0.95,-\h);
				
				\draw[<->]
				(\rG+0.80,-0.05)
				--
				(\rG+0.80,-\h+0.05);
				
				\node[
				right
				] at (\rG+0.80,-0.5*\h)
				{$d_{\mathrm{gap}}$};
				
				% ------------------------------------------------------------
				% Coordinate labels
				% ------------------------------------------------------------
				
				\node[
				right
				] at (\rE+0.30,0)
				{$z=0$};
				
				\node[
				right
				] at (\rG+1.0,-\h)
				{$z=d_{\mathrm{gap}}$};
				
				% ------------------------------------------------------------
				% Labels
				% ------------------------------------------------------------
				
				\node[
				red!75!black,
				align=center,
				left
				] at (-\rE-0.0,0.20)
				{
					Powered electrode\\
					$A_{\mathrm E}=\pi r_{\mathrm E}^2$
				};
				
				\node[
				blue!70!black,
				align=center,
				right
				] at (\rG-1.3,-\h-0.9)
				{
					Grounded electrode\\
					$A_{\mathrm G}=\pi r_{\mathrm G}^2$
				};
				
				\node[
				blue!70!black,
				align=center
				] at (0,-2.8)
				{
					Plasma bulk\\
					(quasineutral)
				};
				
				% ------------------------------------------------------------
				% Radius law
				% ------------------------------------------------------------
				
				\node[
				align=left,
				anchor=west
				] at (-4.8,-1.5)
				{
					$r(z)
					=
					r_{\mathrm E}
					+
					\dfrac{
						r_{\mathrm G}-r_{\mathrm E}
					}{
						d_{\mathrm{gap}}
					}z$
				};
				
				\node[
				align=left,
				anchor=west
				] at (-4.8,-2.25)
				{
					$A(z)=\pi r^2(z)$
				};
				
			\end{tikzpicture}
		}
		\caption{
			Equivalent conical-frustum geometry used for the asymmetric RF discharge. The frustum represents an effective current channel with end areas $A_{\mathrm E}$ and $A_{\mathrm G}$ and axial separation $d_{\mathrm{gap}}$, rather than the detailed reactor geometry.
		}
		\label{fig:equivalent_geometry}
	\end{figure}

	As in the preceding self-consistent model \cite{Mussenbrock2026Minimal}, the reactor is represented by an equivalent conical-frustum current channel. This effective geometry provides a continuous representation of electrode-area asymmetry while retaining closed-form expressions for both the axial bulk response and the quantities entering the stationary global balance. The axial coordinate is $0\le z\le d_{\mathrm{gap}}$, with the powered electrode at $z=0$ and the grounded electrode at $z=d_{\mathrm{gap}}$. The radius varies linearly,
	\begin{gather}
		r(z)=r_{\mathrm E}+\alpha z,
		\qquad
		\alpha=\frac{r_{\mathrm G}-r_{\mathrm E}}{d_{\mathrm{gap}}},
		\label{eq:frustum_radius}
	\end{gather}
	where \(r_{\mathrm E}\) and \(r_{\mathrm G}\) are the effective radii at the powered and grounded boundaries, respectively. The local cross-sectional area is
	\begin{gather}
		A(z)=\pi r^2(z).
		\label{eq:frustum_area}
	\end{gather}
	The end areas are $A_{\mathrm E}=\pi r_{\mathrm E}^2$ and $A_{\mathrm G}=\pi r_{\mathrm G}^2$. Thus, \(A_{\mathrm E}=A_{\mathrm G}\) recovers the symmetric discharge, whereas \(A_{\mathrm E}\ne A_{\mathrm G}\) introduces geometrical asymmetry within the same description. Moreover, the axial geometric factor entering the bulk impedance remains analytic,
	\begin{gather}
		{\int_0^{d_{\mathrm{gap}}}\frac{dz}{A(z)} =\frac{d_{\mathrm{gap}}}{\sqrt{A_{\mathrm E}A_{\mathrm G}}},} 
	\end{gather}
	a relation used below in the bulk momentum balance.
	
	The effective plasma volume used in the stationary global balance is
	\begin{gather}
		V=\frac{d_{\mathrm{gap}}}{3}
		\left(
		A_{\mathrm E}+\sqrt{A_{\mathrm E}A_{\mathrm G}}+A_{\mathrm G}
		\right),
		\label{eq:plasma_volume}
	\end{gather}
	while particle loss to the two electrode-facing sheath boundaries is represented by
	\begin{gather}
		A_{\mathrm{loss}}=A_{\mathrm E}+A_{\mathrm G}.
		\label{eq:loss_area}
	\end{gather}
	Consistent with the quasi-one-dimensional description, particle transport in the reduced geometry is assumed to be predominantly axial, so that losses occur primarily through the two electrode sheaths and lateral-wall losses are neglected.
	
	\subsection{lncosh sheath charge--voltage model with bounded differential elastance}\label{sec:B}
	
	The RF circuit requires a constitutive charge--voltage relation for each boundary sheath. Analytical descriptions of capacitive RF sheaths commonly reduce the boundary region to a nonlinear charge-storage element, with the detailed charge--voltage law depending on the assumed ion-density profile and sheath dynamics \cite{Lieberman1988,GodyakSternberg1990,Metze1986}. A lncosh sheath characteristic was introduced by Ziegler et al. in a nonlinear global model matched to a dual-frequency CCP experiment \cite{ZieglerMussenbrockBrinkmann2008PSST}. In that work, the experimentally reconstructed electrode-sheath charge--voltage relation was quadratic near sheath collapse and nearly linear at larger charge, and an asymptotically linear lncosh form was introduced as a compact representation of this behavior.
	
	For a planar depletion sheath with surface charge density $\sigma=Q/A$ and sheath width $s$, the depletion-sheath charge relation can be written as
	\begin{gather}
		\sigma(s)=e\int_0^s n_{\mathrm i}(z)\,dz,
	\end{gather}
	and therefore
	\begin{gather}
		\frac{ds}{dQ}=\frac{1}{eA n_{\mathrm i}(s)}.
		\label{eq:differential_width_general}
	\end{gather}
	The corresponding differential charge--voltage relation follows from Poisson's equation,
	\begin{gather}
		\frac{dV_{\mathrm S}}{dQ}
		=
		\frac{s}{\varepsilon_0 A}.
		\label{eq:differential_elastance_general}
	\end{gather}
	Hence the sheath characteristic is determined by the dependence of $s$ on $Q$.
	For the transcendental sheath characteristic, Ziegler et al. reconstructed an ion-density profile of the form
	\begin{gather}
		n_{\mathrm i}(z)=\frac{n_0}{1-(z/\overline s)^2},
		\label{eq:ziegler_density_profile}
	\end{gather}
	where \(n_0\) is the density scale and \(\overline s\) the limiting sheath thickness. Evaluating this profile at the instantaneous sheath edge $z=s$ and substituting into Eq.~\eqref{eq:differential_width_general} yields
	\begin{gather}
		\frac{ds}{dQ}
		=
		\frac{1}{e n_0 A}
		\left[
		1-\left(\frac{s}{\overline s}\right)^2
		\right],
	\end{gather}
	which is an autonomous Riccati equation. The reconstructed profile therefore implies a Riccati-type relation between sheath width and charge.
	
	The present model retains this Riccati form as the constitutive closure for both dynamic sheaths, with the small-charge prefactor fixed by the effective sheath-edge density $n_{\mathrm s}$ supplied by the stationary plasma model. The parameter $n_0$ of the reconstructed profile therefore does not enter the present closure. Let $Q_j\ge0$ denote the positive space charge in sheath $j\in\{\mathrm E,\mathrm G\}$, and let $V_{\mathrm S,j}>0$ be the corresponding plasma-to-electrode voltage drop. Using Eq.~\eqref{eq:differential_elastance_general}, the differential sheath elastance is
	\begin{gather}
		S_j(Q_j)
		=
		\frac{dV_{\mathrm S,j}}{dQ_j}
		=
		\frac{1}{C_{\mathrm{diff},j}(Q_j)}
		=
		\frac{s_j(Q_j)}{\varepsilon_0 A_j},
		\label{eq:differential_elastance_width}
	\end{gather}
	where $C_{\mathrm{diff},j}=dQ_j/dV_{\mathrm S,j}$ and $s_j(Q_j)$ is the effective differential sheath width.
	
	For small sheath charge, the standard depletion-sheath relation requires
	\begin{gather}
		s_j(Q_j)
		\simeq
		\frac{Q_j}{e n_{\mathrm s}A_j},
		\qquad
		Q_j\rightarrow0.
		\label{eq:small_charge_width}
	\end{gather}
	Accordingly, the Riccati closure for sheath \(j\) is written as
	\begin{gather}
		\frac{ds_j}{dQ_j}
		=
		\frac{1}{e n_{\mathrm s}A_j}
		\left[
		1-\left(\frac{s_j}{\overline s_j}\right)^2
		\right],
		\qquad
		s_j(0)=0.
		\label{eq:saturating_width_ode}
	\end{gather}
	Here, \(\overline s_j\) denotes the limiting sheath-width scale of sheath j. For $s_j\ll\overline s_j$, Eq.~\eqref{eq:saturating_width_ode} recovers the required linear small-charge width; for $s_j\rightarrow\overline s_j$, $ds_j/dQ_j\to0$. The bounded solution is
	\begin{gather}
		s_j(Q_j)
		=
		\overline s_j
		\tanh\left(\frac{Q_j}{Q_{0,j}}\right),
		\label{eq:tanh_width}
	\end{gather}
	with $Q_{0,j}$ the characteristic charge
	\begin{gather}
		Q_{0,j}
		=
		e n_{\mathrm s}A_j\overline s_j.
		\label{eq:characteristic_charge}
	\end{gather}
	Using Eq.~\eqref{eq:differential_elastance_width} and integrating from zero charge  gives
	\begin{align}
		V_{\mathrm S,j}(Q_j)
		&=
		\frac{\overline s_j}{\varepsilon_0A_j}
		\int_0^{Q_j}
		\tanh\left(\frac{q}{Q_{0,j}}\right)dq
		\\
		&=
		V_{0,j}
		\ln\!\cosh\left(\frac{Q_j}{Q_{0,j}}\right),
		\label{eq:lncosh_sheath}
	\end{align}
	where the corresponding voltage scale is
	\begin{gather}
		V_{0,j}
		=
		\frac{e n_{\mathrm s}\overline s_j^2}{\varepsilon_0}.
		\label{eq:characteristic_voltage}
	\end{gather}
	The corresponding differential elastance is
	\begin{gather}
		S_j(Q_j)
		=
		\frac{\overline s_j}{\varepsilon_0A_j}
		\tanh\left(\frac{Q_j}{Q_{0,j}}\right).
		\label{eq:lncosh_elastance}
	\end{gather}
	
	The relation recovers the quadratic sheath to leading order. Since
	\begin{gather}
		\ln\!\cosh x
		=
		\frac{x^2}{2}-\frac{x^4}{12}+\mathcal O(x^6),
	\end{gather}
	Eq.~\eqref{eq:lncosh_sheath} gives
	\begin{gather}
		V_{\mathrm S,j}
		=
		\frac{Q_j^2}{2e\varepsilon_0 n_{\mathrm s}A_j^2}
		-
		\frac{Q_j^4}{12e^3\varepsilon_0 n_{\mathrm s}^3A_j^4\overline s_j^2}
		+\mathcal O(Q_j^6).
		\label{eq:small_charge_expansion}
	\end{gather}
	The first term is the standard quadratic depletion-sheath result. At large charge,
	\begin{gather}
		S_j(Q_j)
		\rightarrow
		\frac{\overline s_j}{\varepsilon_0A_j},
	\end{gather}
	so that the differential sheath capacitance approaches the finite value $\varepsilon_0A_j/\overline s_j$.
	
	The lncosh relation thus provides a reduced sheath closure that retains the depletion-sheath limit and finite asymptotic elastance while remaining explicit in $V_{\mathrm S,j}(Q_j)$, $S_j(Q_j)$, and $s_j(Q_j)$. In contrast to a purely quadratic sheath law, its charge--voltage characteristic contains higher-order contributions at finite charge.
	
	\subsection{Higher-order sheath nonlinearities in the symmetric two-sheath limit}
	
	The distinction between a purely quadratic sheath approximation and the present lncosh sheath model becomes particularly clear in the symmetric limit, where the two boundary sheaths have identical charge--voltage characteristics. The cancellation of quadratic sheath nonlinearities in this limit and the importance of higher-order terms were identified previously in equivalent-circuit analyses of symmetric and weakly asymmetric CCPs \cite{SchuengelEtAl2015POP}. We use a general Taylor expansion to expose the same structure directly for an arbitrary sheath function $F(Q)$. When conduction currents contribute only weakly to the RF sheath-charge balance, $Q_{\mathrm E}+Q_{\mathrm G}$ varies slowly on the RF timescale and the two sheath charges are approximately complementary. We therefore write
	\begin{gather}
		Q_{\mathrm E}=\overline Q+q,
		\qquad
		Q_{\mathrm G}=\overline Q-q.
		\label{eq:complementary_charges}
	\end{gather}
	Here, $\overline Q=(Q_{\mathrm E}+Q_{\mathrm G})/2$ is the slowly varying common-mode sheath charge and $q=(Q_{\mathrm E}-Q_{\mathrm G})/2$ the RF charge excursion.
	For identical sheath functions $F(Q)$, the voltage contribution entering the series circuit is
	\begin{gather}
		\Delta V_{\mathrm S}
		=
		F(\overline Q+q)-F(\overline Q-q).
		\label{eq:sheath_difference}
	\end{gather}
	Expansion about $\overline Q$ gives
	\begin{gather}
		\Delta V_{\mathrm S}
		=
		2F'(\overline Q)q
		+
		\frac{1}{3}F'''(\overline Q)q^3
		+
		\frac{1}{60}F^{(5)}(\overline Q)q^5
		+\cdots .
		\label{eq:symmetry_expansion}
	\end{gather}
	All even powers cancel by symmetry. For a purely quadratic law $F(Q)=aQ^2$, all derivatives of order three and higher vanish and
	\begin{gather}
		\Delta V_{\mathrm S}=4a\overline Q q,
		\label{eq:quadratic_cancellation}
	\end{gather}
	which is exactly linear in the RF charge excursion $q$. Thus, the quadratic nonlinearity of the individual sheaths cancels from their differential series voltage. For the lncosh law, by contrast, higher odd derivatives are generally nonzero. The two identical sheaths still cancel all even contributions, but cubic and higher odd terms remain. The earlier cubic-sheath argument of Ref.~\cite{SchuengelEtAl2015POP} is therefore recovered as the lowest-order case: the relevant requirement is the presence of nonvanishing higher odd derivatives of the sheath charge--voltage characteristic about the operating point.
	
\subsection{Characteristic sheath scale}

The sheath law of Sec.~\ref{sec:B} contains the characteristic scale $\overline s_j$, which is determined from the slowly responding ion sheath. The effective sheath-edge density entering both the Bohm flux and the nonlinear sheath charge--voltage relation is retained from the stationary global-model closure \cite{LiebermanLichtenberg2005,Mussenbrock2026Minimal} and written as
\begin{gather}
	n_{\mathrm s}=h n_{\mathrm e},
\end{gather}
with
\begin{gather}
	h
	=
	0.86
	\left(
	3+\frac{d_{\mathrm{gap}}}{2\lambda_{\mathrm i}}
	\right)^{-1/2},
	\qquad
	\lambda_{\mathrm i}
	=
	\frac{1}{n_{\mathrm g}\sigma_{\mathrm i}}.
	\label{eq:h}
\end{gather}
For argon, the ion--neutral momentum-transfer cross section entering the mean free path $\lambda_{\mathrm i}$ is taken from Phelps \cite{Phelps1991}. The same effective sheath-edge density $n_{\mathrm s}$ is used for both dynamic sheaths.

Since the ions respond predominantly to the RF-averaged sheath field, $\overline s_j$ is obtained from the RF-averaged sheath voltage. For the low-pressure conditions considered here, the collisionless Child--Langmuir space-charge law \cite{Child1911,Langmuir1913} together with the Bohm ion flux, in the standard plasma-sheath form summarized in Ref.~\cite{LiebermanLichtenberg2005}, provides the algebraic closure
\begin{gather}
	J_{\mathrm i}
	=
	\frac{4}{9}\varepsilon_0
	\sqrt{\frac{2e}{m_{\mathrm i}}}
	\frac{\overline V_{\mathrm S,j}^{3/2}}{\overline s_j^2},
	\label{eq:child_langmuir_scale}
\end{gather}
with the Bohm ion current density
\begin{gather}
	J_{\mathrm i}=e n_{\mathrm s}v_{\mathrm B},
	\qquad
	v_{\mathrm B}=\sqrt{\frac{eT_{\mathrm e}}{m_{\mathrm i}}},
	\label{eq:bohm_current_density}
\end{gather}
and
\begin{gather}
	\overline V_{\mathrm S,j}
	=
	\left\langle V_{\mathrm S,j}\right\rangle_{\mathrm{RF}}.
\end{gather}
Hence
\begin{gather}
	\overline s_j
	=
	\left[
	\frac{4\varepsilon_0}{9J_{\mathrm i}}
	\sqrt{\frac{2e}{m_{\mathrm i}}}
	\overline V_{\mathrm S,j}^{3/2}
	\right]^{1/2}.
	\label{eq:sbar_closure}
\end{gather}
The two characteristic scales are iterated independently because the
RF-averaged voltage drops can differ in an asymmetric discharge.
Equation~\eqref{eq:sbar_closure} fixes only the slow sheath scale; the instantaneous charge--voltage dynamics remain governed by the lncosh relation of Eq.~\eqref{eq:lncosh_sheath}.
At pressures for which ion--neutral collisions dominate sheath transport,
Eq.~\eqref{eq:sbar_closure} must be replaced by an appropriate collisional
sheath-scale closure, without changing the structure of the four RF
dynamical equations.
	
	\subsection{Ion and electron currents at both sheaths}
	
	The ion and electron conduction currents are included at both electrodes using the usual Bohm-flux and Maxwellian electron-flux expressions \cite{LiebermanLichtenberg2005}. With one common sheath-edge density $n_{\mathrm s}$, the Bohm ion-current magnitudes are
	\begin{gather}
		I_{\mathrm i,j}
		=
		e n_{\mathrm s}v_{\mathrm B}A_j,
		\qquad
		j\in\{\mathrm E,\mathrm G\}.
		\label{eq:ion_currents_two_sheaths}
	\end{gather}
	The corresponding Maxwellian electron saturation currents are
	\begin{gather}
		I_{\mathrm e0,j}
		=
		\frac14e n_{\mathrm s}\overline v_{\mathrm e}A_j,
		\qquad
		\overline v_{\mathrm e}
		=
		\sqrt{\frac{8eT_{\mathrm e}}{\pi m_{\mathrm e}}},
	\end{gather}
	and the instantaneous electron currents are
	\begin{gather}
		I_{\mathrm e,j}
		=
		I_{\mathrm e0,j}
		\exp\left(-\frac{V_{\mathrm S,j}}{T_{\mathrm e}}\right).
		\label{eq:electron_currents_two_sheaths}
	\end{gather}
	Here $T_{\mathrm e}$ is expressed in electronvolts and the sheath voltages in volts.
	
	\subsection{Plasma bulk}
	
	The quasineutral bulk is described by the reduced electron momentum balance, assuming a spatially uniform electron density and that the RF conduction current is carried predominantly by electrons, while pressure-gradient and convective terms are neglected,
	\begin{gather}
		\frac{\partial\vec j}{\partial t}
		=
		\frac{e^2n_{\mathrm e}}{m_{\mathrm e}}\vec E
		-
		\nu_{\mathrm{eff}}\vec j.
	\end{gather}
	Current continuity in the frustum gives $j(z,t)=I(t)/A(z)$. Integration over the current path yields
	\begin{gather}
		V_{\mathrm p}(t)
		=
		L_{\mathrm p}\frac{dI}{dt}
		+
		R_{\mathrm p}I,
		\label{eq:bulk_voltage}
	\end{gather}
	with
	\begin{gather}
		L_{\mathrm p}
		=
		\frac{m_{\mathrm e}d_{\mathrm{gap}}}
		{e^2n_{\mathrm e}\sqrt{A_{\mathrm E}A_{\mathrm G}}},
		\qquad
		R_{\mathrm p}=\nu_{\mathrm{eff}}L_{\mathrm p}.
		\label{eq:bulk_elements}
	\end{gather}
	The effective damping frequency is retained from the reduced model of Ref.~\cite{Mussenbrock2026Minimal},
	\begin{gather}
		\nu_{\mathrm{eff}}
		=
		\nu_{\mathrm m}
		+
		\frac{\overline v_{\mathrm e}}{d_{\mathrm{gap}}},
		\label{eq:effective_damping}
	\end{gather}
	where the second term represents collisionless sheath heating phenomenologically as an effective resistive contribution. This is a circuit-level representation of the long-studied collisionless interaction of electrons with oscillating RF sheaths \cite{Godyak1972,GoeddeLichtenbergLieberman1988,LiebermanGodyak1998,Turner2009Review}; it should not be interpreted as a microscopic momentum-transfer collision frequency.
	
	\subsection{External circuit, multi-harmonic excitation, and dynamical equations}
	
The powered electrode is connected to the RF source through a blocking capacitor $C_{\mathrm B}$. Denoting the powered-electrode voltage relative to ground by $V_{\mathrm E}(t)$, we define
\begin{gather}
	V_{\mathrm B}(t)
	=
	V_{\mathrm{RF}}(t)-V_{\mathrm E}(t),
\end{gather}
so that
\begin{gather}
	\frac{dV_{\mathrm B}}{dt}
	=
	\frac{I}{C_{\mathrm B}}.
	\label{eq:dVb}
\end{gather}
	The applied voltage is written in the general zero-mean form
	\begin{gather}
		V_{\mathrm{RF}}(t)
		=
		\widehat V_{\mathrm{RF}}
		\sum_{n\ge1}
		a_n
		\cos\left(n\omega_{\mathrm{RF}}t+\varphi_n\right).
		\label{eq:multiharmonic_voltage}
	\end{gather}
	A single frequency corresponds to $a_1=1$ and $a_{n>1}=0$. A two-harmonic waveform with a controllable relative phase is the canonical realization of the electrical asymmetry effect \cite{HeilEtAl2008JPD,DonkoEtAl2009JPD,CzarnetzkiEtAl2009JPCS,SchulzeEtAl2009JPD}.

	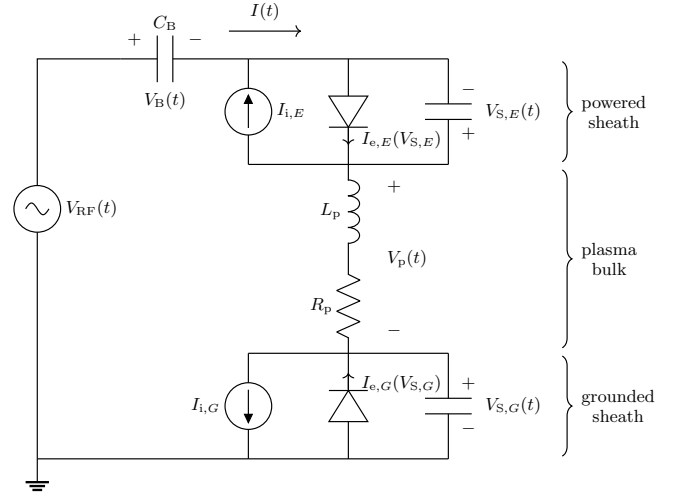
\begin{figure}[t]
		\centering
		\resizebox{\columnwidth}{!}{%
			\begin{circuitikz}[
				american,
				font=\small,
				line width=0.6pt
				]
				
				\ctikzset{
					bipoles/thickness=1
				}
				
				% ------------------------------------------------------------
				% Coordinates
				% ------------------------------------------------------------
				
				\coordinate (GND)  at (0,0);
				\coordinate (SRCB) at (0,1.8);
				\coordinate (SRCT) at (0,7.2);
				
				\coordinate (CBL) at (1.5,7.2);
				\coordinate (CBR) at (3.1,7.2);
				
				% Powered sheath
				\coordinate (SHTL) at (3.8,7.2);
				\coordinate (SHTR) at (7.4,7.2);
				\coordinate (SHBL) at (3.8,5.3);
				\coordinate (SHBR) at (7.4,5.3);
				
				% Plasma bulk
				\coordinate (SHB) at (5.6,5.3);
				\coordinate (LPB) at (5.6,3.6);
				\coordinate (RPB) at (5.6,1.9);
				
				% Grounded sheath
				\coordinate (GHTL) at (3.8,1.9);
				\coordinate (GHTR) at (7.4,1.9);
				\coordinate (GHBL) at (3.8,0.0);
				\coordinate (GHBR) at (7.4,0.0);
				
				% ------------------------------------------------------------
				% RF source and return
				% ------------------------------------------------------------
				
				\draw
				(GND)
				node[ground]{}
				to[short] (SRCB)
				to[sV,l_={$V_{\mathrm{RF}}(t)$}] (SRCT)
				to[short] (CBL);
				
				% ------------------------------------------------------------
				% Blocking capacitor
				% ------------------------------------------------------------
				
				\draw
				(CBL)
				to[C,l^={$C_{\mathrm B}$}] (CBR)
				to[short] (SHTL);
				
				\node at (1.75,7.55) {$+$};
				\node at (2.85,7.55) {$-$};
				\node at (2.30,6.45) {$V_{\mathrm B}(t)$};
				
				% ------------------------------------------------------------
				% Current arrow
				% ------------------------------------------------------------
				
				\draw[->]
				(3.45,7.70)
				--
				(4.75,7.70);
				
				\node at (4.10,8.05) {$I(t)$};
				
				% ------------------------------------------------------------
				% Powered sheath
				% ------------------------------------------------------------
				
				\draw
				(SHTL)
				--
				(SHTR);
				
				\draw
				(SHBL)
				--
				(SHBR);
				
				% Ion current
				\draw
				(SHBL)
				to[isource,l_={$I_{\mathrm i,E}$}]
				(SHTL);
				
				% Electron current
				\draw
				(5.6,7.2)
				to[D]
				(5.6,5.3);
				
				\draw[->]
				(5.6,5.95)
				--
				(5.6,5.65);
				
				\node[right] at (5.72,5.75)
				{$I_{\mathrm e,E}(V_{\mathrm S,E})$};
				
				% Nonlinear sheath element
				\draw
				(SHTR)
				to[C]
				(SHBR);
				
				\node[right] at (7.52,6.65) {$-$};
				\node[right] at (7.52,5.85) {$+$};
				
				\node[right] at (7.95,6.25)
				{$V_{\mathrm S,E}(t)$};
				
				% ------------------------------------------------------------
				% Plasma bulk
				% ------------------------------------------------------------
				
				\draw
				(SHB)
				to[L,l_={$L_{\mathrm p}$}]
				(LPB)
				to[R,l_={$R_{\mathrm p}$}]
				(RPB);
				
				\node[right] at (6.18,4.90) {$+$};
				\node[right] at (6.18,2.30) {$-$};
				\node[right] at (6.18,3.60) {$V_{\mathrm p}(t)$};
				
				% ------------------------------------------------------------
				% Grounded sheath
				% ------------------------------------------------------------
				
				\draw
				(GHTL)
				--
				(GHTR);
				
				\draw
				(GHBL)
				--
				(GHBR);
				
				% Ion current
				\draw
				(GHTL)
				to[isource,l_={$I_{\mathrm i,G}$}]
				(GHBL);
				
				% Electron current
				\draw
				(5.6,0.0)
				to[D]
				(5.6,1.9);
				
				\draw[->]
				(5.6,1.25)
				--
				(5.6,1.55);
				
				\node[right] at (5.72,1.35)
				{$I_{\mathrm e,G}(V_{\mathrm S,G})$};
				
				% Nonlinear sheath element
				\draw
				(GHTR)
				to[C]
				(GHBR);
				
				\node[right] at (7.52,1.35) {$+$};
				\node[right] at (7.52,0.55) {$-$};
				
				\node[right] at (7.95,0.95)
				{$V_{\mathrm S,G}(t)$};
				
				% ------------------------------------------------------------
				% Return to ground
				% ------------------------------------------------------------
				
				\draw
				(GHBL)
				to[short]
				(GND);
				
				% ------------------------------------------------------------
				% Braces and labels
				% ------------------------------------------------------------
				
				% Powered sheath
				\draw[
				decorate,
				decoration={
					brace,
					amplitude=5pt
				}
				]
				(9.45,7.15)
				--
				(9.45,5.35);
				
				\node[
				right,
				align=center
				] at (9.68,6.25)
				{powered\\sheath};
				
				% Plasma bulk
				\draw[
				decorate,
				decoration={
					brace,
					amplitude=5pt
				}
				]
				(9.45,5.20)
				--
				(9.45,2.00);
				
				\node[
				right,
				align=center
				] at (9.68,3.60)
				{plasma\\bulk};
				
				% Grounded sheath
				\draw[
				decorate,
				decoration={
					brace,
					amplitude=5pt
				}
				]
				(9.45,1.85)
				--
				(9.45,0.05);
				
				\node[
				right,
				align=center
				] at (9.68,0.95)
				{grounded\\sheath};
				
			\end{circuitikz}
		}
		\caption{
			Equivalent circuit of the asymmetric CCRF discharge. Both boundary sheaths are represented by nonlinear charge-storage elements together with the corresponding ion and electron conduction currents. The plasma bulk is represented by $L_{\mathrm p}$ and $R_{\mathrm p}$.
		}
		\label{fig:equivalent_circuit}
	\end{figure}
	
	Positive current is defined from the powered electrode toward the grounded electrode. Charge conservation at the two sheaths then gives
	\begin{align}
		\frac{dQ_{\mathrm E}}{dt}
		&=
		I_{\mathrm e,E}-I_{\mathrm i,E}-I,
		\label{eq:dQE}
		\\
		\frac{dQ_{\mathrm G}}{dt}
		&=
		I+I_{\mathrm e,G}-I_{\mathrm i,G}.
		\label{eq:dQG}
	\end{align}
	With $V_{\mathrm S,E}$ and $V_{\mathrm S,G}$ evaluated from Eq.~\eqref{eq:lncosh_sheath}, Kirchhoff's voltage law yields
	\begin{gather}
		\frac{dI}{dt}
		=
		\frac{1}{L_{\mathrm p}}
		\left[
		V_{\mathrm{RF}}(t)
		-
		V_{\mathrm B}
		+
		V_{\mathrm S,E}
		-
		V_{\mathrm S,G}
		-
		R_{\mathrm p}I
		\right].
		\label{eq:dI}
	\end{gather}
	Equations~\eqref{eq:dVb}, \eqref{eq:dQE}, \eqref{eq:dQG}, and \eqref{eq:dI} form the four-dimensional nonlinear RF subsystem. Both sheath charges are restricted to $Q_j\ge0$, with $Q_j=0$ corresponding to sheath collapse.
	
	In the stationary periodic state,
	\begin{gather}
		\left\langle I\right\rangle_{\mathrm{RF}}=0,
	\end{gather}
	and the dc self-bias is
	\begin{gather}
		V_{\mathrm{SB}}
		=
		-\left\langle V_{\mathrm B}\right\rangle_{\mathrm{RF}}.
		\label{eq:self_bias}
	\end{gather}
	
\section{Stationary plasma closure for argon}
	
To close the nonlinear RF subsystem self-consistently, the slow plasma state is determined using the standard global-model construction \cite{LiebermanLichtenberg2005}, retained in the same reduced form as in Ref.~\cite{Mussenbrock2026Minimal}. For an electropositive argon discharge, the sheath-edge density is $n_{\mathrm s}=hn_{\mathrm e}$, with $h$ given by Eq.~\eqref{eq:h}. The neutral density is
	\begin{gather}
		n_{\mathrm g}=\frac{p}{k_{\mathrm B}T_{\mathrm g}}.
	\end{gather}
	The stationary particle balance is
	\begin{gather}
		n_{\rm e}n_{\mathrm g}K_{\mathrm{iz}}(T_{\mathrm e})V
		=
		h n_{\mathrm e}v_{\mathrm B}A_{\mathrm{loss}},
	\end{gather}
	which gives
	\begin{gather}
		n_{\mathrm g}K_{\mathrm{iz}}(T_{\mathrm e})
		=
		hv_{\mathrm B}(T_{\mathrm e})
		\frac{A_{\mathrm{loss}}}{V}.
		\label{eq:particle_balance}
	\end{gather}
	For given pressure, gas temperature, geometry, and $h$, this relation determines $T_{\mathrm e}$. The cancellation of \(n_{\mathrm e}\) reflects the standard electropositive global-model result that the stationary particle balance fixes the electron temperature independently of the plasma density \cite{LiebermanLichtenberg2005}.
	
	The cycle-averaged power transferred to the electron population is
	\begin{gather}
		P_{\mathrm{abs}}
		=
		R_{\mathrm p}
		\left\langle I^2\right\rangle_{\mathrm{RF}}.
		\label{eq:absorbed_power}
	\end{gather}
	The stationary energy balance is written as
	\begin{gather}
		P_{\mathrm{abs}}
		=
		e h n_{\mathrm e}v_{\mathrm B}A_{\mathrm{loss}}
		\mathcal E_{\mathrm{eff}}(T_{\mathrm e}),
	\end{gather}
	so that
	\begin{gather}
		n_{\mathrm e}
		=
		\frac{P_{\mathrm{abs}}}
		{e h v_{\mathrm B}A_{\mathrm{loss}}\mathcal E_{\mathrm{eff}}(T_{\mathrm e})}.
		\label{eq:density_closure}
	\end{gather}
	
	The reduced Maxwellian rate coefficients are retained unchanged. Here and below, $T_{\mathrm e}$ is expressed in electronvolts, with
	\begin{gather}
		v_{\mathrm B}(T_{\mathrm e})
		=
		\sqrt{\frac{eT_{\mathrm e}}{m_{\mathrm i}}},
		\quad
		\overline v_{\mathrm e}(T_{\mathrm e})
		=
		\sqrt{\frac{8eT_{\mathrm e}}{\pi m_{\mathrm e}}},
	\end{gather}
	Electron-impact ionization is represented by \cite{LiebermanLichtenberg2005}
	\begin{gather}
		K_{\mathrm{iz}}(T_{\mathrm e})
		=
		2.34\times10^{-14}
		T_{\mathrm e}^{0.59}
		\exp\left(-\frac{17.44}{T_{\mathrm e}}\right)
		\quad\mathrm{m^3\,s^{-1}},
		\label{eq:Kiz}
	\end{gather}
	excitation by
	\begin{gather}
		K_{\mathrm{ex}}(T_{\mathrm e})
		=
		2.48\times10^{-14}
		T_{\mathrm e}^{0.33}
		\exp\left(-\frac{12.78}{T_{\mathrm e}}\right)
		\quad\mathrm{m^3\,s^{-1}},
		\label{eq:Kex}
	\end{gather}
	and momentum transfer by
	\begin{multline}
		K_{\mathrm m}(T_{\mathrm e})
		=
		2.336\times10^{-14}T_{\mathrm e}^{1.609}
		\\
		\times
		\exp\left[
		0.0618(\ln T_{\mathrm e})^2
		-
		0.1171(\ln T_{\mathrm e})^3
		\right]
		\quad\mathrm{m^3\,s^{-1}}.
		\label{eq:Km}
	\end{multline}
	The momentum-transfer data are consistent with the argon cross-section data available through the Phelps/LXCat compilations \cite{YamabeBuckmanPhelps1983,PitchfordEtAl2017}. The same analytical fit is used for the elastic energy-transfer coefficient $K_{\mathrm{el}}$. The effective electron energy loss per created electron--ion pair is
	\begin{gather}
		\mathcal E_{\mathrm{eff}}
		=
		15.76
		+
		\frac{K_{\mathrm{ex}}}{K_{\mathrm{iz}}}12.14
		+
		\frac{K_{\mathrm{el}}}{K_{\mathrm{iz}}}
		\frac{3m_{\mathrm e}}{m_{\mathrm i}}T_{\mathrm e}
		+
		2T_{\mathrm e}.
		\label{eq:energy_loss}
	\end{gather}
	
\section{Self-consistent numerical coupling}\label{sec:coupling}

The model contains three nested levels. For prescribed $n_{\mathrm e}$ and $T_{\mathrm e}$, the periodic solution of Eqs.~\eqref{eq:dVb}, \eqref{eq:dQE}, \eqref{eq:dQG}, and \eqref{eq:dI} is obtained by a shooting method. The phase-zero RF state
\begin{gather}
	\mathbf{y}_0
	=
	\left(
	Q_{\mathrm E},
	Q_{\mathrm G},
	V_{\mathrm B},
	I
	\right)_{t=0}
\end{gather}
is determined such that
\begin{gather}
	\mathbf{y}(T_{\mathrm{RF}};\mathbf{y}_0)
	-
	\mathbf{y}_0
	=
	\mathbf{0}.
\end{gather}
The shooting residual is normalized by characteristic scales of the four state variables, and convergence is assumed when its maximum absolute component falls below $2\times10^{-6}$. The sheath charges are represented by logarithmic variables within the shooting procedure, which guarantees positive trial values. For continuation within the self-consistent iteration and parameter scans, the previously converged RF state is reused as the initial estimate for the next shooting solve.

For the resulting periodic orbit, the RF-averaged sheath voltages determine the characteristic scales $\overline s_{\mathrm E}$ and $\overline s_{\mathrm G}$ through Eq.~\eqref{eq:sbar_closure}. These slow sheath closures are iterated to convergence while the periodic RF state is reused as the initial estimate of the next iteration.

The resulting absorbed power is then inserted into the stationary electron-energy balance, Eq.~\eqref{eq:density_closure}, while the particle balance, Eq.~\eqref{eq:particle_balance}, updates $T_{\mathrm e}$. The new plasma state changes $n_{\mathrm s}$, $L_{\mathrm p}$, $R_{\mathrm p}$, the conduction currents, and the two nonlinear sheath characteristics. This closes the self-consistent feedback loop between plasma sustainment and the nonlinear RF response.

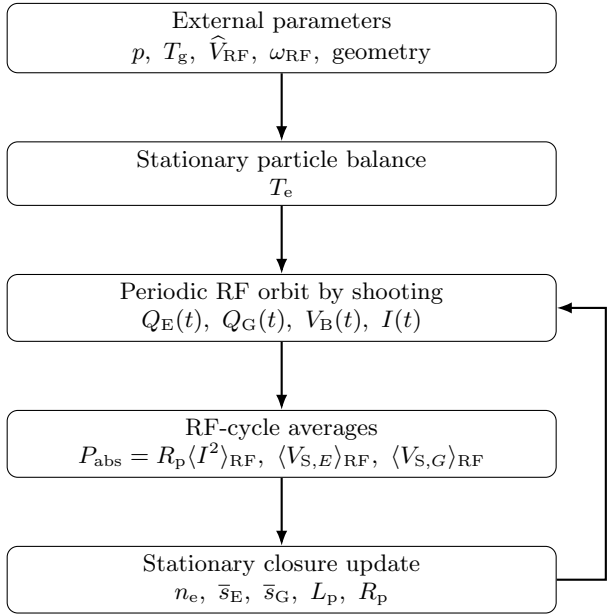
\begin{figure}[t]
	\centering
	\begin{tikzpicture}[
		box/.style={
			draw,
			rounded corners,
			align=center,
			minimum width=0.84\columnwidth,
			minimum height=0.82cm
		},
		arrow/.style={
			-{Latex[length=2mm]},
			thick
		},
		node distance=0.9cm
		]
		
		\node[box] (inputs)
		{
			External parameters\\
			$p,\ T_{\mathrm g},\
			\widehat V_{\mathrm{RF}},\
			\omega_{\mathrm{RF}},\
			\text{geometry}$
		};
		
		\node[box,below=of inputs] (pb)
		{
			Stationary particle balance\\
			$T_{\mathrm e}$
		};
		
		\node[box,below=of pb] (rf)
		{
			Periodic RF orbit by shooting\\
			$Q_{\mathrm E}(t),\
			Q_{\mathrm G}(t),\
			V_{\mathrm B}(t),\
			I(t)$
		};
		
		\node[box,below=of rf] (avg)
		{
			RF-cycle averages\\
			$P_{\mathrm{abs}}
			=
			R_{\mathrm p}
			\langle I^2\rangle_{\mathrm{RF}},\
			\langle V_{\mathrm S,E}\rangle_{\mathrm{RF}},\
			\langle V_{\mathrm S,G}\rangle_{\mathrm{RF}}$
		};
		
		\node[box,below=of avg] (sc)
		{
			Stationary closure update\\
			$n_{\mathrm e},\
			\overline s_{\mathrm E},\
			\overline s_{\mathrm G},\
			L_{\mathrm p},\
			R_{\mathrm p}$
		};
		
		\draw[arrow] (inputs) -- (pb);
		\draw[arrow] (pb) -- (rf);
		\draw[arrow] (rf) -- (avg);
		\draw[arrow] (avg) -- (sc);
		
		\draw[arrow]
		(sc.east)
		--
		++(0.65,0)
		|-
		(rf.east);
		
	\end{tikzpicture}
	
	\caption{
		Self-consistent coupling between the stationary plasma closure and
		the periodic RF model. For fixed external parameters, the stationary
		particle balance determines $T_{\mathrm e}$. The periodic RF orbit is
		then obtained by solving the one-period shooting condition for the
		four-dimensional RF state. The resulting absorbed power and
		RF-averaged sheath voltages update the stationary plasma state and the
		effective circuit parameters before the RF solution is repeated until
		convergence.
	}
	\label{fig:coupling}
\end{figure}

The self-consistent density update is under-relaxed,
\begin{gather}
	n_{\mathrm e}^{(k+1)}
	=
	(1-\alpha)n_{\mathrm e}^{(k)}
	+
	\alpha n_{\mathrm e,tar}^{(k+1)},
\end{gather}
with $\alpha=0.3$ in the present calculations. Convergence of the outer plasma-state iteration is assumed when the maximum relative change of $T_{\mathrm e}$ and $n_{\mathrm e}$ falls below $10^{-5}$. The characteristic sheath scales are updated with an under-relaxation factor of $0.5$ and are considered converged when their maximum relative change falls below $2\times10^{-4}$. These relaxation parameters affect convergence only and do not introduce additional physical timescales into the stationary model.

Each evaluation of the shooting residual requires integration over one RF period. The RF equations are integrated using the adaptive DOP853 method with a relative tolerance of $10^{-9}$ and absolute tolerances of $10^{-15}\,\mathrm{C}$ for $Q_{\mathrm E}$ and $Q_{\mathrm G}$, $10^{-8}\,\mathrm{V}$ for $V_{\mathrm B}$, and $10^{-8}\,\mathrm{A}$ for $I$. After convergence of the shooting problem, one final RF period is integrated on the output grid used for RF averaging and spectral analysis. The physical constraint $Q_j\ge0$ is additionally enforced in the RF equations: at $Q_j=0$, any derivative that would drive the sheath charge to negative values is set to zero. If the shooting iteration fails to reach the prescribed periodicity tolerance, a cycle-by-cycle integration procedure is retained as a numerical fallback.
	
	\section{Characteristic plasma-series-resonance scale}
	
	The nonlinear sheath--bulk system possesses a characteristic
	plasma-series-resonance (PSR) frequency associated with the inertia of
	the bulk electrons and the differential capacitances of the two
	sheaths \cite{Godyak1986,AnnaratoneAllenKu1995,
		KuAnnaratoneAllen1998a,KuAnnaratoneAllen1998b}.
	For the present lncosh sheath model, the instantaneous differential
	elastances are
	\begin{align}
		S_{\mathrm E}(t)
		&=
		\frac{\overline s_{\mathrm E}}
		{\varepsilon_0 A_{\mathrm E}}
		\tanh\left(
		\frac{Q_{\mathrm E}(t)}{Q_{0,\mathrm E}}
		\right),
		\\
		S_{\mathrm G}(t)
		&=
		\frac{\overline s_{\mathrm G}}
		{\varepsilon_0 A_{\mathrm G}}
		\tanh\left(
		\frac{Q_{\mathrm G}(t)}{Q_{0,\mathrm G}}
		\right).
	\end{align}
	Since the two sheaths are traversed in series, their differential
	elastances add,
	\begin{gather}
		S_{\mathrm{sh}}(t)
		=
		S_{\mathrm E}(t)
		+
		S_{\mathrm G}(t).
		\label{eq:total_sheath_elastance}
	\end{gather}
	
	For a small high-frequency perturbation of the periodic RF state, the
	two sheath elastances act in series with the blocking capacitor.
	Neglecting damping for the purpose of identifying the characteristic
	frequency scale gives
	\begin{gather}
		\omega_{\mathrm{PSR}}(t)
		=
		\sqrt{
			\frac{
				S_{\mathrm{sh}}(t)
				+
				C_{\mathrm B}^{-1}
			}{
				L_{\mathrm p}
			}
		}.
		\label{eq:local_psr}
	\end{gather}
	Equivalently, the corresponding harmonic number relative to the applied
	RF frequency is
	\begin{gather}
		k_{\mathrm{PSR}}(t)
		=
		\frac{\omega_{\mathrm{PSR}}(t)}
		{\omega_{\mathrm{RF}}}.
		\label{eq:local_psr_harmonic}
	\end{gather}
	
	Because the sheath charges and therefore the differential elastances
	vary throughout the RF cycle, the nonlinear system does not possess a
	single fixed LC resonance frequency. Equation~\eqref{eq:local_psr}
	should therefore be interpreted as a time-dependent characteristic PSR
	scale. It is used below to relate enhanced higher harmonics in the
	calculated current spectra to the characteristic frequency range of the
	sheath--bulk series resonance. 
	
The interpretation of $\omega_{\mathrm{PSR}}(t)$ as a local resonance scale implicitly assumes a sufficient separation between the fast PSR dynamics and the slower variation of the underlying RF state. Determining the parameter range over which this quasi-static picture remains valid, and where the separation of time scales eventually breaks down, constitutes an interesting question for future work.
	
	\section{Results}
	
	We use a geometrically symmetric discharge driven by a single frequency voltage waveform as the
	base case. The externally prescribed operating and geometric parameters
	are summarized in Table~\ref{tab:base_case}. Unless stated otherwise,
	these parameters are kept fixed throughout the following calculations.
All plasma quantities, including the electron temperature
$T_{\mathrm e}$, electron density $n_{\mathrm e}$, the characteristic sheath scales
$\overline s_{\mathrm E}$ and $\overline s_{\mathrm G}$, and the
effective bulk parameters $L_{\mathrm p}$ and $R_{\mathrm p}$, are
determined self-consistently based on Sec.~\ref{sec:coupling}.
	
	\begin{table}[t]
		\centering
		\caption{
			Base-case parameters used in the calculations. The base case
			corresponds to a geometrically symmetric discharge under
			monofrequent excitation.
		}
		\label{tab:base_case}
		\begin{tabular}{l l l}
			\hline
			Quantity & Symbol & Value \\
			\hline
			
			Powered-electrode radius
			& $r_{\mathrm E}$
			& $10\,\mathrm{cm}$ \\
			
			Grounded-electrode radius
			& $r_{\mathrm G}$
			& $10\,\mathrm{cm}$ \\
			
			Axial separation
			& $d_{\mathrm{gap}}$
			& $5.0\,\mathrm{cm}$ \\
			
			Blocking capacitance
			& $C_{\mathrm B}$
			& $10\,\mathrm{nF}$ \\
			
			RF frequency
			& $f_{\mathrm{RF}}$
			& $13.56\,\mathrm{MHz}$ \\
			
			RF voltage amplitude
			& $\widehat V_{\mathrm{RF}}$
			& $250\,\mathrm{V}$ \\
			
			Gas pressure
			& $p$
			& $1.0\,\mathrm{Pa}$ \\
			
			Neutral-gas temperature
			& $T_{\mathrm g}$
			& $300\,\mathrm{K}$ \\
			
			\hline
		\end{tabular}
	\end{table}

	\subsection{Geometrical asymmetry under single frequency excitation}
	
	We first investigate geometrical symmetry breaking while retaining a
	purely monofrequent excitation,
	\begin{gather}
		V_{\mathrm{RF}}(t)
		=
		\widehat V_{\mathrm{RF}}
		\cos(\omega_{\mathrm{RF}}t).
	\end{gather}
	The grounded-electrode radius is kept fixed at
	$r_{\mathrm G}=10\,\mathrm{cm}$, 
while the powered-electrode radius \(r_{\mathrm E}\) is varied from \(2\) to \(10\,\mathrm{cm}\).
	
	All remaining parameters are kept at their base-case values. This way, the
	scan continuously connects a strongly geometrically asymmetric
	configuration to the symmetric limit
	$r_{\mathrm E}=r_{\mathrm G}$.
	
\begin{figure}[t]
		\centering
		\includegraphics[width=0.9\linewidth]
		{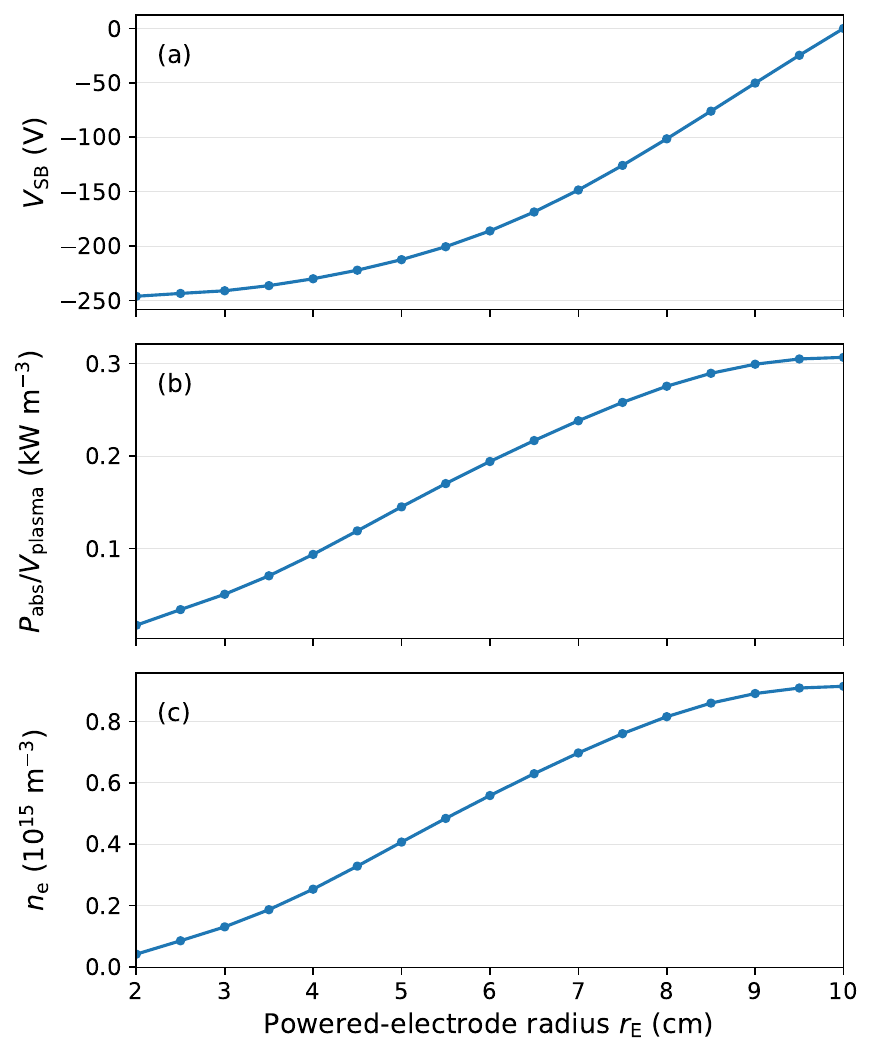}
		\caption{
			Self-consistent operating point as a function of geometrical
			asymmetry under monofrequent excitation. The grounded-electrode
			radius is fixed at $r_{\mathrm G}=10\,\mathrm{cm}$ and the
			powered-electrode radius is varied from $2$ to $10\,\mathrm{cm}$.
			(a) dc self-bias voltage,
			(b) volume-averaged absorbed power density
			$P_{\mathrm{abs}}/V_{\mathrm{plasma}}$, and
			(c) electron density $n_{\mathrm e}$.
			The symmetric limit is reached at
			$r_{\mathrm E}=r_{\mathrm G}=10\,\mathrm{cm}$.
		}
		\label{fig:geometry_scan_overview}
	\end{figure}

	Figure~\ref{fig:geometry_scan_overview} shows the resulting
	self-consistent operating point as a function of the powered-electrode
	radius. The dc self-bias voltage is negative for
	$r_{\mathrm E}<r_{\mathrm G}$ and decreases in magnitude continuously
	as the two electrode radii approach each other. In the symmetric limit,
	the self-bias vanishes, as required by the geometrical symmetry of the
	discharge. For the strongest asymmetry considered here,
	$r_{\mathrm E}=2\,\mathrm{cm}$, the magnitude of the self-bias reaches
	approximately $240\,\mathrm{V}$.
	
	The geometrical variation also affects the power absorbed by the plasma.
	Figure~\ref{fig:geometry_scan_overview}(b) shows the volume-averaged
	absorbed power density
	\begin{gather}
		p_{\mathrm{abs}}
		=
		\frac{P_{\mathrm{abs}}}{V_{\mathrm{plasma}}},
	\end{gather}
	which increases as the discharge approaches the symmetric
	configuration. This behavior should not be identified directly with the
	strength of the nonlinear RF response. Within the present lumped model,
	the absorbed power is determined by
	\begin{gather}
		P_{\mathrm{abs}}
		=
		R_{\mathrm p}
		\left\langle I^2\right\rangle_{\mathrm{RF}},
	\end{gather}
	where both the current amplitude and the self-consistent bulk resistance
	change with geometry.
	
	The corresponding electron density is shown in
	Fig.~\ref{fig:geometry_scan_overview}(c). Its dependence on geometry
	results self-consistently from the coupled particle and energy balances.
	Using the stationary particle balance, the absorbed power density can
	be written as
	\begin{gather}
		\frac{P_{\mathrm{abs}}}{V_{\mathrm{plasma}}}
		=
		e n_{\mathrm e} n_{\mathrm g}
		K_{\mathrm{iz}}(T_{\mathrm e})
		\mathcal E_{\mathrm{eff}}(T_{\mathrm e}),
	\end{gather}
	so that the similar trends of absorbed power density and electron
	density reflect their coupling through the stationary plasma closure.
	The electron temperature itself changes with the geometrical loss
	factor $A_{\mathrm{loss}}/V_{\mathrm{plasma}}$.
	
	\begin{figure}[t]
		\centering
		\includegraphics[width=\linewidth]
		{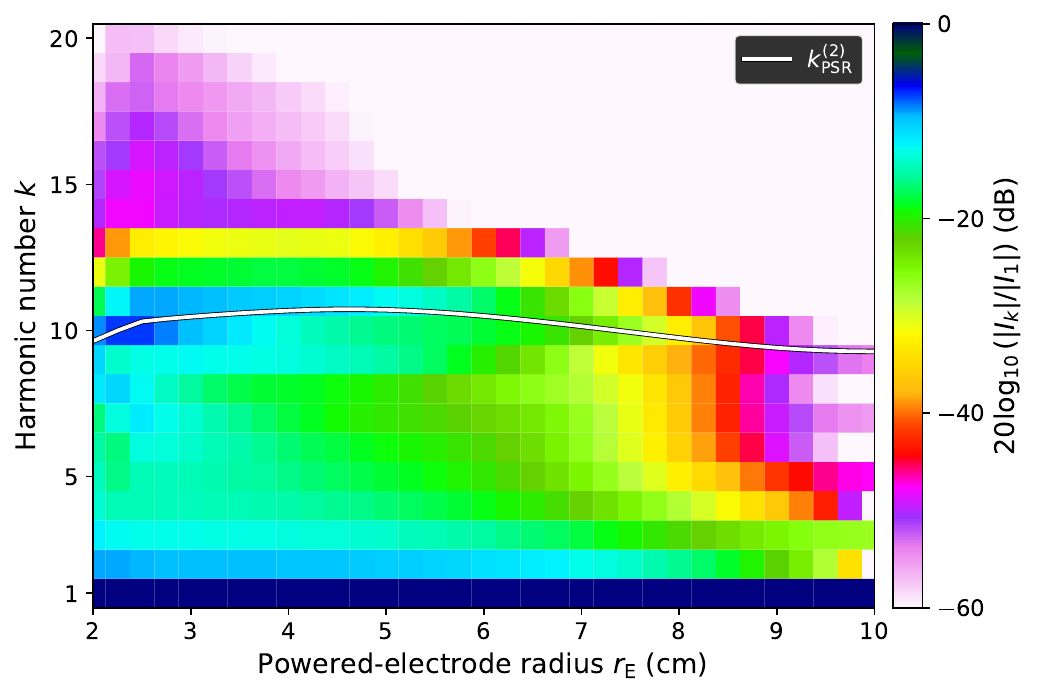}
		\caption{
			Normalized current-harmonic spectrum during the geometrical-
			asymmetry scan of Fig.~\ref{fig:geometry_scan_overview}.
			The color scale shows
			$20\log_{10}(|I_k|/|I_1|)$ for
			$k=1,\ldots,20$.
			The white line indicates the moment-weighted characteristic
			plasma-series-resonance scale $k_{\mathrm{PSR}}^{(2)}$.
			In the symmetric limit, the half-wave symmetry of the current
			eliminates all even harmonics, whereas geometrical symmetry
			breaking allows both even and odd harmonics and gives rise to an
			increasingly pronounced higher-harmonic response.
		}
		\label{fig:geometry_scan_harmonics}
	\end{figure}

	The change in the self-consistent operating point is accompanied by a
	pronounced change in the nonlinear RF dynamics.
	Figure~\ref{fig:geometry_scan_harmonics} shows the current-harmonic
	spectrum over the same geometrical scan. For each geometry, the
	harmonic amplitudes are normalized to the fundamental component and
	displayed as
	\begin{gather}
		20\log_{10}
		\left(
		\frac{|I_k|}{|I_1|}
		\right).
	\end{gather}
	
	A particularly clear limiting behavior is obtained for the symmetric
	discharge. In this case, the periodic current obeys the half-wave
	symmetry
	\begin{gather}
		I\left(t+\frac{T_{\mathrm{RF}}}{2}\right)
		=
		-I(t),
	\end{gather}
	and therefore contains only odd harmonics. All even current harmonics
	vanish exactly within the symmetric model, apart from numerical
	residuals. This harmonic selection rule is consistent with the
	symmetry behavior observed in kinetic PIC/MCC simulations
	\cite{NoesgesMussenbrock2025POP}. As the powered-electrode radius is
	reduced, the geometrical symmetry is broken, the half-wave symmetry is
	lost, and even harmonics become allowed. At the same time, the spectrum
	develops an increasingly pronounced higher-harmonic structure.
	
	The white line in Fig.~\ref{fig:geometry_scan_harmonics} indicates the
	moment-weighted characteristic plasma-series-resonance scale,
	\begin{gather}
		k_{\mathrm{PSR}}^{(2)}
		=
		\frac{1}{\omega_{\mathrm{RF}}}
		\sqrt{
			\frac{
				S_{\mathrm{eff}}^{(2)}
				+
				C_{\mathrm B}^{-1}
			}{
				L_{\mathrm p}
			}
		},
	\end{gather}
	with
	\begin{gather}
		S_{\mathrm{eff}}^{(2)}
		=
		\frac{
			\left\langle S_{\mathrm{sh}}^3\right\rangle_{\mathrm{RF}}
		}{
			\left\langle S_{\mathrm{sh}}^2\right\rangle_{\mathrm{RF}}
		},
		\qquad
		S_{\mathrm{sh}}
		=
		S_{\mathrm E}+S_{\mathrm G}.
	\end{gather}
	With increasing geometrical asymmetry, the characteristic PSR scale
	shifts toward higher harmonic numbers and follows the spectral region
	in which the higher current harmonics become particularly pronounced.
	
	The quantity \(k_{\mathrm{PSR}}^{(2)}\) is determined entirely by the differential sheath response and the bulk electron inertia and is independent of the calculated current spectrum. The fact that the enhanced current harmonics follow this scale therefore reflects the underlying sheath--bulk dynamics: nonlinear sheath charging generates high-frequency current components, while the series combination of the sheath elastances and the inductive plasma bulk preferentially enhances components near the PSR scale. As the geometrical asymmetry increases, the sheath operating points and their differential elastances change, shifting this characteristic scale through the harmonic spectrum.

	\subsection{Interplay of geometrical and electrical asymmetry}
	
	We next consider the combined effect of geometrical and electrical
	symmetry breaking. The grounded-electrode radius is kept fixed at
	$r_{\mathrm G}=10\,\mathrm{cm}$, while the powered-electrode radius
	$r_{\mathrm E}$ and the phase angle $\theta$ of a two-harmonic driving
	voltage are varied simultaneously. The applied voltage is written as
	\begin{gather}
		V_{\mathrm{RF}}(t)
		=
		\frac{\widehat V_{\mathrm{RF}}}{2}
		\left[
		\cos(\omega_{\mathrm{RF}}t+\theta)
		+
		\cos(2\omega_{\mathrm{RF}}t)
		\right].
	\end{gather}
	The powered-electrode radius is varied between
	$4$ and $10\,\mathrm{cm}$, while the phase angle is scanned from
	$0^\circ$ to $90^\circ$. All remaining parameters are kept at their
	base-case values.
	
	This two-dimensional parameter scan allows geometrical and electrical
	asymmetry to be treated within the same framework. The geometrically
	symmetric case is recovered at $r_{\mathrm E}=r_{\mathrm G}$, whereas
	$r_{\mathrm E}<r_{\mathrm G}$ introduces geometrical asymmetry. At
	each geometry, variation of $\theta$ changes the electrical asymmetry
	of the applied waveform.
	
\begin{figure}[t]
		\centering
		\includegraphics[width=\linewidth]
		{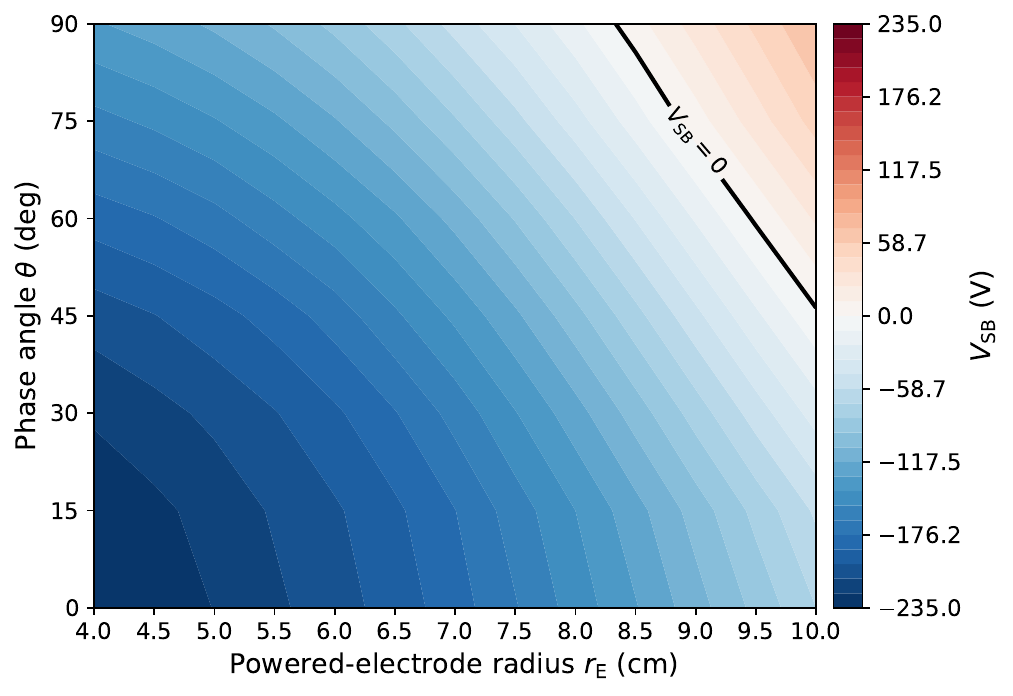}
		\caption{
			Dc self-bias voltage $V_{\mathrm{SB}}$ as a function of
			powered-electrode radius $r_{\mathrm E}$ and phase angle
			$\theta$ for fixed $r_{\mathrm G}=10\,\mathrm{cm}$.
			The black contour marks $V_{\mathrm{SB}}=0$.
		}
		\label{fig:interplay_asymmetry}
	\end{figure}

	Figure~\ref{fig:interplay_asymmetry} shows the resulting dc self-bias
	voltage as a function of $r_{\mathrm E}$ and $\theta$. The contour
	defined by
	\begin{gather}
		V_{\mathrm{SB}}=0
	\end{gather}
	marks operating points for which the dc self-bias is compensated.
	Along the geometrically symmetric boundary, variation of the phase
	angle generates a finite dc self-bias purely through electrical
	symmetry breaking. For geometrically asymmetric configurations, the
	electrically generated contribution can either enhance or compensate
	the geometrically induced self-bias.
	
	Most notably, the zero-bias contour extends into the region
	$r_{\mathrm E}<r_{\mathrm G}$. Hence, for a finite range of geometrical
	asymmetries, the dc self-bias can be compensated by an appropriate
	choice of the phase angle. For stronger geometrical asymmetry, the
	available phase range is no longer sufficient to reach
	$V_{\mathrm{SB}}=0$. This behavior is consistent with previous PIC/MCC simulations and
	analytical modeling of the electrical compensation of geometrical
	asymmetry in capacitively coupled plasmas
	\cite{SchulzeEtAl2011APL,SaikiaEtAl2018JAP}.
	
\begin{figure}[t]
		\centering
		\includegraphics[width=\linewidth]
		{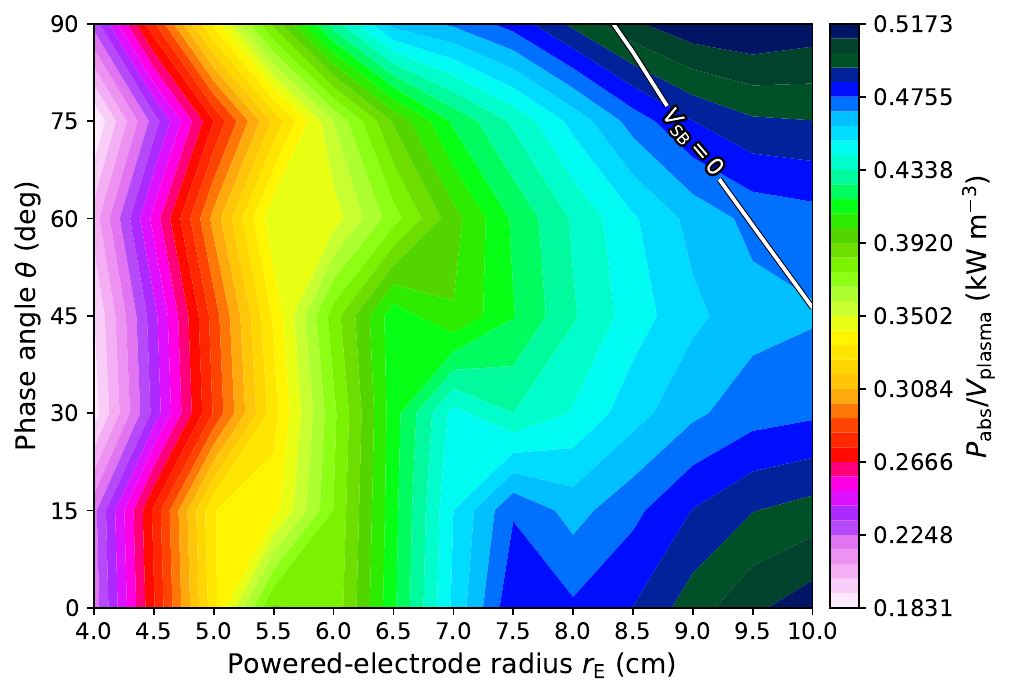}
		\caption{
			Volume-averaged absorbed power density
			$P_{\mathrm{abs}}/V_{\mathrm{plasma}}$ over the same
			$(r_{\mathrm E},\theta)$ parameter space as in
			Fig.~\ref{fig:interplay_asymmetry}.
			The white line with black outline marks the contour
			$V_{\mathrm{SB}}=0$.
		}
		\label{fig:interplay_power}
	\end{figure}

	The same parameter variation also changes the global plasma operating
	point. Figure~\ref{fig:interplay_power} shows the corresponding
	volume-averaged absorbed power density,
	\begin{gather}
		p_{\mathrm{abs}}
		=
		\frac{P_{\mathrm{abs}}}{V_{\mathrm{plasma}}}.
	\end{gather}
	In contrast to the dc self-bias, which is strongly affected by both
	geometry and phase, the absorbed power density exhibits a dominant
	dependence on the powered-electrode radius, with a weaker but
	systematic phase dependence. This dominant geometrical dependence is consistent with the stationary
	closure. Variation of $r_{\mathrm E}$ changes the geometrical loss
	factor $A_{\mathrm{loss}}/V_{\mathrm{plasma}}$ and therefore the
	particle-balance electron temperature, while the simultaneous change
	of $A_{\mathrm{bulk,eff}}=\sqrt{A_{\mathrm E}A_{\mathrm G}}$ modifies
	the bulk inductance and resistance. Consequently, configurations with
	the same dc self-bias need not share the same global power balance.
	Over the parameter range considered here,
	the absorbed power density varies approximately between
	$0.11$ and $0.52\,\mathrm{kW\,m^{-3}}$.
	
	The $V_{\mathrm{SB}}=0$ contour is superimposed in
	Fig.~\ref{fig:interplay_power}. Its position demonstrates that
	compensation of the dc self-bias does not correspond to a fixed global
	plasma operating point. Different zero-bias configurations can exhibit
	different absorbed power densities even though their time-averaged dc
	bias vanishes.
	
	The dc self-bias and the absorbed power density characterize only
	time-averaged aspects of the discharge. To quantify the nonlinear
	spectral response separately from the harmonics directly contained in
	the applied voltage, we introduce the generated spectral fraction
	\begin{gather}
		\eta_{\mathrm{NL}}
		=
		\frac{
			\displaystyle
			\sum_{\substack{k=1\\k\notin\mathcal D}}^{K}
			|I_k|^2
		}{
			\displaystyle
			\sum_{k=1}^{K}
			|I_k|^2
		},
	\end{gather}
	where $\mathcal D$ denotes the set of directly driven harmonics and
	$K$ is the largest harmonic retained in the evaluation. For the
	present two-harmonic excitation,
	\begin{gather}
		\mathcal D=\{1,2\},
		\qquad
		K=20,
	\end{gather}
	so that all spectral components with $k\geq3$ contribute to the
	numerator.
	
\begin{figure}[t]
		\centering
		\includegraphics[width=\linewidth]
		{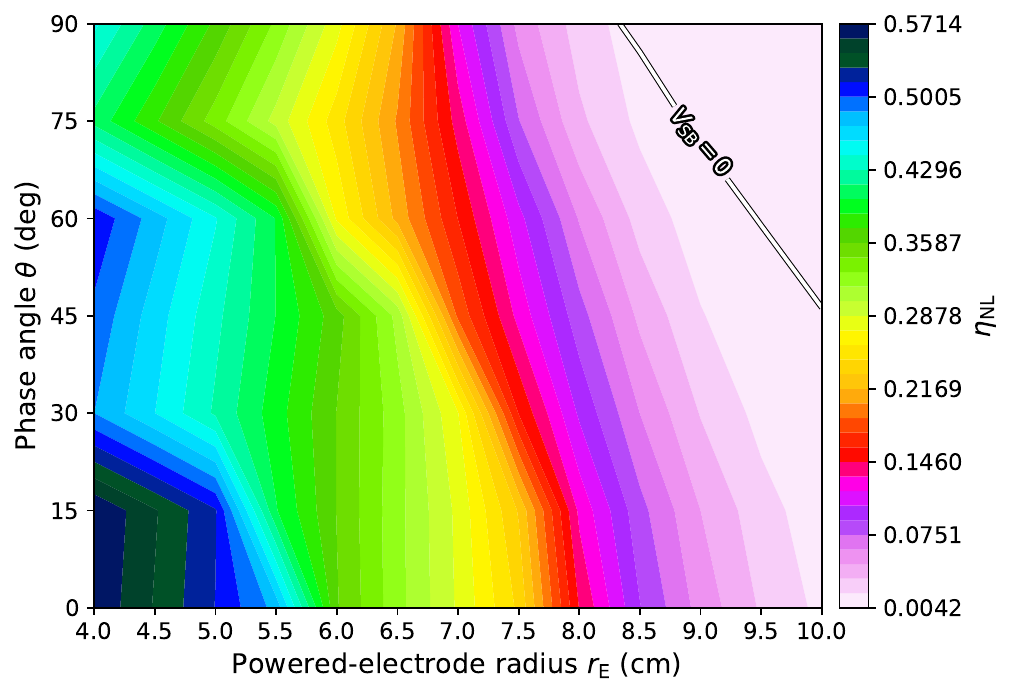}
		\caption{
			Generated spectral fraction $\eta_{\mathrm{NL}}$ as a function
			of powered-electrode radius $r_{\mathrm E}$ and phase angle
			$\theta$. Harmonics directly present in the applied voltage are
			excluded from the numerator. For the present excitation,
			$\mathcal D=\{1,2\}$ and $K=20$. The white contour with black
			outline marks $V_{\mathrm{SB}}=0$.
		}
		\label{fig:interplay_nonlinear_fraction}
	\end{figure}

	Figure~\ref{fig:interplay_nonlinear_fraction} shows
	$\eta_{\mathrm{NL}}$ over the same parameter space. A pronounced
	localization of the nonlinear spectral response is obtained. For
	strong geometrical asymmetry, a substantial fraction of the current
	spectrum is transferred to harmonics that are absent from the applied
	voltage. The generated spectral fraction reaches values close to
	$0.6$ in the present scan.
	
	Toward the nearly symmetric configurations,
	$\eta_{\mathrm{NL}}$ decreases strongly. In particular, the generated
	spectral fraction is small in the vicinity of the zero-self-bias
	contour. Thus, within the parameter range considered here, compensation
	of the dc self-bias is accompanied by a marked reduction of the
	generated spectral content. Conversely, a vanishing dc self-bias is
	not used here as a general criterion for dynamical symmetry; the
	spectral measure provides an independent characterization of the
	nonlinear RF response.
	
	Finally, we examine whether the moment-weighted PSR scale introduced above also describes the spectral position of independently detected high-frequency structure in the generated current spectrum. Rather than identifying the largest generated harmonic, we search for local maxima of the spectrum after excluding the harmonics directly contained in the applied voltage. This distinction is important because a low-order generated component may carry more spectral power than a weaker but
	well-localized high-frequency peak.
	
	For the present two-harmonic excitation, $\mathcal D=\{1,2\}$ and the generated spectrum therefore starts at $k=3$. Local maxima are identified in the generated spectral amplitudes $|I_k|$ for $k\leq K$, with $K=20$. The first generated harmonic is treated as the lower boundary of this spectral range rather than as a local maximum, since it has no generated-harmonic neighbor on its low-frequency side. This prevents a monotonically decreasing low-order spectral contribution from being interpreted automatically as a separate high-frequency structure.
	
	For each local maximum, we determine its spectral prominence $p_k$ relative to the surrounding spectral minima. To suppress numerically insignificant extrema, only peaks satisfying
	\begin{gather}
		p_k
		\geq
		\epsilon
		\max_{\substack{1\leq q\leq K\\q\notin\mathcal D}}
		|I_q|
	\end{gather}
	are retained, with $\epsilon=0.01$ in the present analysis. Among the accepted local maxima, the most distinctly localized spectral feature is defined by
	\begin{gather}
		k_{\mathrm{HF}}
		=
		\underset{k\in\mathcal P}{\operatorname{arg\,max}}\,
		p_k,
	\end{gather}
	where $\mathcal P$ denotes the set of accepted local maxima. If no significant local maximum is present, no value of $k_{\mathrm{HF}}$ is assigned to that operating point.
	
\begin{figure}[t]
		\centering
		\includegraphics[width=0.9\linewidth]
		{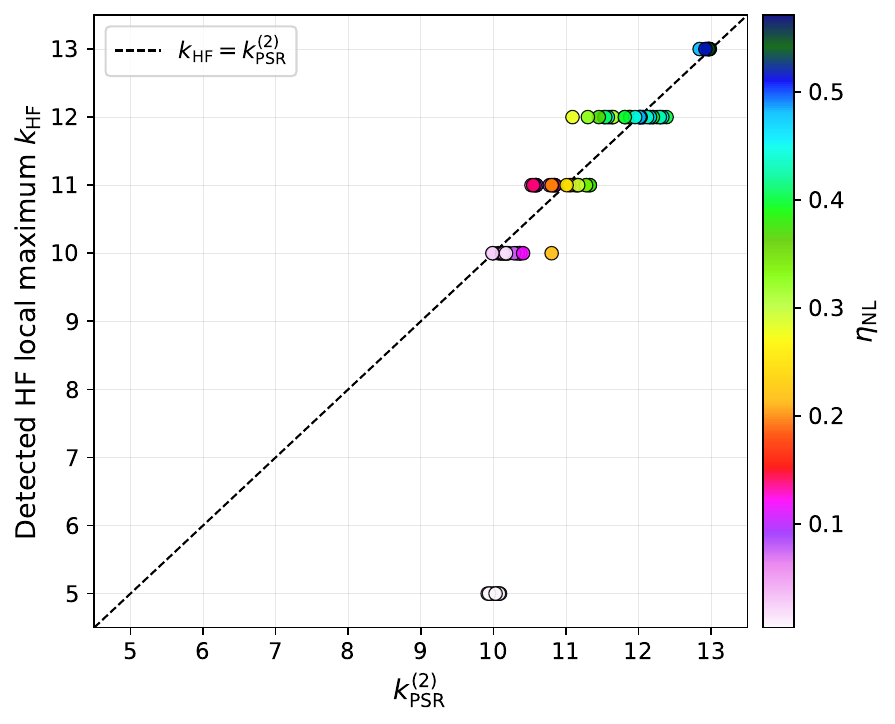}
		\caption{
			Detected high-frequency local maximum $k_{\mathrm{HF}}$ versus the moment-weighted characteristic plasma-series-resonance scale $k_{\mathrm{PSR}}^{(2)}$ for the two-dimensional $(r_{\mathrm E},\theta)$ scan. The local maximum is determined independently from the generated current spectrum without using $k_{\mathrm{PSR}}^{(2)}$ in the peak-detection procedure. The color denotes the generated spectral fraction $\eta_{\mathrm{NL}}$. The dashed diagonal indicates $k_{\mathrm{HF}}=k_{\mathrm{PSR}}^{(2)}$.
		}
		\label{fig:psr_peak_comparison}
	\end{figure}

	Importantly, this procedure uses only the calculated current spectrum. The independently determined quantity $k_{\mathrm{PSR}}^{(2)}$ does not enter the peak-detection algorithm. The comparison in Fig.~\ref{fig:psr_peak_comparison} therefore provides a non-circular test of whether a localized high-frequency spectral feature appears near the characteristic PSR scale.
	
	Figure~\ref{fig:psr_peak_comparison} compares the detected $k_{\mathrm{HF}}$ with the moment-weighted characteristic PSR scale $k_{\mathrm{PSR}}^{(2)}$. The color of each point denotes the corresponding generated spectral fraction $\eta_{\mathrm{NL}}$. For operating points with an appreciable nonlinear spectral response, the detected local maxima cluster close to
	\begin{gather}
		k_{\mathrm{HF}}
		=
		k_{\mathrm{PSR}}^{(2)}.
	\end{gather}
	In particular, the pronounced high-frequency structures around harmonic numbers $k\simeq10$--$13$ follow the independently calculated
	PSR scale closely. Since $k_{\mathrm{HF}}$ is restricted to integer harmonic numbers whereas $k_{\mathrm{PSR}}^{(2)}$ is a continuous characteristic frequency scale, the resulting points naturally form discrete horizontal bands rather than lying exactly on the diagonal.
	
	The few larger deviations occur for operating points with very small $\eta_{\mathrm{NL}}$. In these cases, a weak low-order local maximum can satisfy the peak-detection criterion even though no pronounced high-frequency nonlinear response is present. Such points therefore carry little information about PSR-dominated spectral generation. By contrast, when the generated spectral fraction becomes substantial, the independently detected high-frequency maximum follows $k_{\mathrm{PSR}}^{(2)}$ systematically.

\section{Discussion and conclusions}

The principal extension introduced here is not simply the addition of a
second sheath degree of freedom. A second dynamic sheath combined with
the same purely quadratic charge--voltage relation would strongly
reduce the leading nonlinear contribution in the symmetric limit.
A sheath characteristic that retains higher-order nonlinear terms is
therefore structurally required if a compact two-sheath circuit model
is to describe symmetric, geometrically asymmetric, and electrically
asymmetric discharges within the same formulation. The lncosh sheath
law provides one compact realization of this requirement. It follows
from the Riccati closure for the differential sheath width, recovers the
quadratic depletion-sheath relation in the small-charge limit, and
approaches a finite asymptotic differential elastance at large charge.
Its particular analytic form is not unique; the essential feature is
that the nonlinear sheath voltage, differential elastance, and effective
sheath width remain explicit functions of the sheath charge.

The parameter studies demonstrate that the same reduced formulation
consistently describes geometrical and electrical symmetry breaking,
dc self-bias formation and compensation, changes of the global plasma
state, and nonlinear harmonic generation. In particular, compensation
of the dc self-bias does not imply dynamical equivalence of the two
boundary regions: along the zero-bias contour, the absorbed power
density and the nonlinear spectral response can still vary with
geometry and excitation phase. The dc self-bias is a time-averaged
circuit property, whereas the nonlinear RF dynamics depend on the
instantaneous charge--voltage characteristics of both sheaths.

The comparison between independently detected high-frequency spectral
structure and the moment-weighted scale
$k_{\mathrm{PSR}}^{(2)}$ provides a more specific test of the nonlinear
circuit dynamics. When a pronounced high-frequency response develops,
the detected spectral maximum lies close to the independently
calculated characteristic PSR scale. For operating points with only
weak generated spectral content, local spectral maxima may still occur,
but their position alone does not provide evidence for a PSR-dominated
response. Accordingly, $k_{\mathrm{PSR}}^{(2)}$ should not be regarded
as an exact eigenfrequency of the periodically driven nonlinear
circuit. Rather, it is a characteristic scale that emphasizes phases
of large total sheath elastance and provides a compact diagnostic of
PSR-like dynamics within the present reduced model.

Several limitations remain. The lncosh sheath law is a reduced
constitutive relation rather than a kinetic sheath solution. The
characteristic sheath scales $\overline s_j$ are determined from a
collisionless Child--Langmuir/Bohm closure and would require a different
slow-scale relation when collisional ion transport becomes important.
Such a replacement would not require modification of the
Riccati/lncosh RF structure itself. In addition, the plasma bulk is
represented by a single inductive--resistive element with an effective
length fixed at $d_{\mathrm{gap}}$, so that $L_{\mathrm p}$ remains
constant within an RF period. PIC and equivalent-circuit studies have
shown that a time-dependent bulk electron plasma frequency can provide
an additional nonlinearity relevant to PSR self-excitation in symmetric
or weakly asymmetric discharges
\cite{SchuengelEtAl2015POP,WilczekEtAl2016POP}. The present model should
therefore be understood as an isolation of the nonlinear two-sheath
circuit dynamics rather than as a complete kinetic description of all
possible excitation mechanisms.

Within these limits, the model provides a self-consistent and
computationally inexpensive description that connects nonlinear RF
dynamics to a stationary global plasma state across symmetric,
geometrically asymmetric, and electrically asymmetric configurations.
It therefore constitutes an intermediate description between purely
analytical circuit models and spatially resolved kinetic simulations
and is particularly suited to systematic parameter studies.

	\begin{acknowledgments}
		This work was funded by the Deutsche Forschungsgemeinschaft (DFG, German Research Foundation) under research grant MU 2332/12-1 (project no. 534102992).
	\end{acknowledgments}
	
	\section*{Data Availability}
	The data that support the findings of this study are available from the corresponding author upon reasonable request.
	
	\bibliography{M-004-NonlinearTwoSheathCCP-References}
	
\end{document}